\documentclass[12pt]{article}
\usepackage[a4paper,textwidth=490.0pt,textheight=703.1pt]{geometry}
\usepackage{url}
\usepackage{booktabs}

\usepackage{epsfig} %
\usepackage{float} %
\usepackage{amssymb} %
\usepackage{amsmath} %
\usepackage{verbatim} %
\usepackage{cite} %
\usepackage{color} %
\usepackage{eufrak}
\usepackage[english]{babel}
\usepackage[latin1]{inputenc}
\usepackage[T1]{fontenc}
\usepackage{ae}

\newcommand{\GeV}{{\rm GeV}}

\newcommand{\beq}{\begin{equation}}
\newcommand{\eeq}{\end{equation}}
\newcommand{\bea}{\begin{eqnarray}}
\newcommand{\eea}{\end{eqnarray}}

\allowdisplaybreaks[4]

\begin{document} 
\setlength{\baselineskip}{0.515cm}

\sloppy 
\thispagestyle{empty} 
\begin{flushleft} DESY 21--013
\\ 
DO--TH 21/04
\\ 
TTP21--004
\\ 
SAGEX--21--03
\\ 
February 2021 \end{flushleft}

\mbox{} \vspace*{\fill} \begin{center}

{\LARGE\bf The QED Initial State Corrections to}

\vspace*{2mm} {\LARGE\bf the Forward-Backward Asymmetry of}

\vspace*{2mm} {\LARGE \bf \boldmath $e^+e^- \rightarrow \gamma^*/{Z^{0}}^*$ to Higher Orders}

\vspace{3cm} \large {\large J.~Bl\"umlein$^a$, A.~De~Freitas$^a$, and K.~Sch\"onwald$^{b}$ }

\vspace{1.cm} \normalsize {\it $^a$~Deutsches Elektronen--Synchrotron, DESY,}\\ {\it Platanenallee 6, D--15738 Zeuthen, 
Germany}

\vspace*{2mm} {\it $^b$~Institut f\"ur Theoretische Teilchenphysik,\\ Karlsruher Institut f\"ur Technologie (KIT) D-76128 
Karlsruhe, Germany}


\end{center} 
\normalsize 
\vspace{\fill} 
\begin{abstract} 
\noindent 
The QED initial state corrections are calculated to the forward--backward asymmetry for 
$e^+e^- \rightarrow \gamma^*/{Z^{0}}^*$ in the leading logarithmic approximation to 
$O(\alpha^6 L^6)$ extending the known corrections up to $O(\alpha^2 L^2)$ in analytic 
form. We use the method of massive on-shell operator matrix elements and present the 
radiators both in Mellin-$N$ and momentum fraction $z$-space. Numerical results are 
presented for various energies around the $Z$-peak by also including energy cuts. These 
corrections are of relevance for the precision measurements at the FCC$\_$ee. 
\end{abstract}

\vspace*{1cm}
\begin{center}
{\sf
Dedicated to the Memory of Tini Veltman, who made it possible to probe the \\ Standard Model
at high precision.}
\end{center}

\vspace*{\fill} \noindent
\newpage

\section{Introduction} 
\label{sec:1}

\vspace*{1mm} 
\noindent 
The measurement of the forward--backward asymmetry for the process $e^+e^- \rightarrow \gamma^*/Z^*$ 
provides an excellent possibility to determine the running fine structure constant $\alpha_{\rm QED}(s) = 4 \pi a(s)$ near $s = 
M_Z^2$ at 
high precision. At the planned future $e^+e^-$ facilities which operate at high energy and at large luminosity, like the ILC, 
CLIC \cite{ILC,Aihara:2019gcq,Mnich:2019,CLIC}, the FCC\_ee \cite{FCCEE}, and also muon colliders \cite{Delahaye:2019omf}, if 
operating in the vicinity of the $Z$-peak, one will obtain highly precise data. Any measurement based on these data needs
a theoretical description of even higher precision \cite{Veltman:2000xp}. Recently, higher order inclusive corrections 
were calculated for $e^+e^- \rightarrow \gamma^*/Z^*$ to $O(\alpha_s^6 L^5)$ up to the first three orders in $L$ at the 
respective 
order in $\alpha$ in Refs.~\cite{Blumlein:2019srk,Blumlein:2019pqb,Blumlein:2020jrf,Ablinger:2020qvo} confirming the results 
to $O(\alpha^2)$ in Ref.~\cite{Blumlein:2011mi} and correcting Refs.~\cite{Berends:1987ab,Kniehl:1988id}, where $L = 
\ln(s/m_e^2)$. This will lead to a 
change of the analysis codes {\tt TOPAZ} \cite{Montagna:1998kp,BP} and {\tt  ZFITTER} \cite{ZFITTER} and may require a 
re-analysis of the data taken at LEP \cite{ALEPH:2005ab}.

The first order QED initial state radiative (ISR) corrections to the forward--backward asymmetry have been 
mutually calculated, 
see \cite{Bohm:1989pb,Beenakker:1989km,BP,QED,Bardin:1989cw,Bardin:1990de,Riemann:1989ca}. Furthermore, the initial-final 
state interference and final state corrections are known at this order, cf.~\cite{BP} for a survey. Furthermore, electroweak 
\cite{AFB:EW,Bardin:1990fu} and QCD corrections \cite{AFB:QCD} have also been calculated. Starting at $O(\alpha^2)$ also the 
contributions to 
the leading order series of $O(\alpha^k L^k),~k \geq 1$, with $L = \ln(s/m_e^2)$\footnote{The leading order contributions to 
the direct terms were obtain in \cite{LL}.} receive besides the inclusive contribution another one, related to the angular 
structure, being present in all sub-leading terms as well. Yet the $O(\alpha^k L^k)$ terms are universal, since they do not 
depend on the process-dependent Wilson coefficients, cf.~\cite{Ablinger:2020qvo}. The inclusive terms were computed in 
\cite{Ablinger:2020qvo} and include besides the anomalous dimensions and massive OMEs also the inclusive massless Wilson 
coefficients up to $O(\alpha^2)$ \cite{Hamberg:1990np,Harlander:2002wh}.

In the present paper we will calculate the angular dependent leading logarithmic contributions to the radiators to $O(\alpha^6 
L^6)$ as a first specific contribution which emerges for the forward--backward asymmetry. To 2nd order, these corrections were 
obtained in \cite{Beenakker:1989km}. The corresponding radiators can be represented by iterated integrals over the alphabet of 
the harmonic polylogarithms \cite{Remiddi:1999ew} and cyclotomic harmonic polylogarithms \cite{Ablinger:2011te} for cyclotomy 
{\sf c = 4}. We also determine efficient representations for these quantities allowing a fast numerical analysis and present 
the 
corresponding corrections for the forward--backward asymmetry in the vicinity of the $Z$-resonance. These corrections, unlike 
the inclusive ones, do not lead to distribution--valued radiators.

The paper is organized as follows. In Section~\ref{sec:2} we calculate the higher order QED initial state corrections to the 
forward--backward asymmetry. Here we use the packages {\tt Sigma} \cite{SIG1,SIG2} and {\tt HarmonicSums} 
\cite{HARMSU, Ablinger:2011te,Vermaseren:1998uu,Blumlein:1998if,Remiddi:1999ew,Blumlein:2009ta}. 
The leading logarithmic radiators of $H_{FB}(z)$ to $O(\alpha^6 L^6)$ of the angular-dependent 
terms are presented in Section~\ref{sec:3}.  We also derive the 
expansions of the radiators in the regions $z \rightarrow 0$ and $z \rightarrow 1$. Numerical and phenomenological results are 
presented in Section~\ref{sec:4}. In Appendix~\ref{sec:A} we give a brief account of the aforementioned radiators in Mellin 
$N$-space. 
The radiators are given in computer-readable form in an attachment to this paper.

\vspace*{-0.5cm}
\section{The Forward-Backward Asymmetry around the \boldmath{$Z$}-Peak} 
\label{sec:2}

\vspace*{1mm} \noindent The forward-backward asymmetry is formed out of the partial cross sections integrating over the angle 
$\theta$ for the forward and backward hemispheres separately,
\begin{eqnarray}
	\sigma_F = 2\pi \int\limits_{0}^{1} {\rm d} \cos(\theta) \frac{{\rm d}\sigma}{{\rm d} \Omega},~~~~~~ \sigma_B = 2\pi 
	\int\limits_{-1}^{0} {\rm d} \cos(\theta) \frac{{\rm d}\sigma}{{\rm d} \Omega} \,.
\end{eqnarray}
The angle $\theta$ is defined between the incoming electron $e^-$ and the outgoing muon $\mu^-$ from $\gamma^*/Z^*$ 
decay.
The forward-backward asymmetry is defined by
\begin{eqnarray}
	A_{\text{FB}}(s) &=& \frac{ \sigma_F(s) - \sigma_B(s)}{ \sigma_{T}(s) }, \end{eqnarray}
with $\sigma_T(s) = \sigma_F(s) + \sigma_B(s)$. At Born level this reduces to \cite{BDJ}
\begin{eqnarray}
	\sigma_{FB}^{(0)}(s) &=& \sigma_F^{(0)}(s) - \sigma_B^{(0)}(s) = \frac{\pi\alpha^2}{s} N_{C,f} \left( 
	1-\frac{4m_f^2}{s} \right) G_3(s) \,, 
\\ 
\sigma_{T}^{(0)}(s)  &=& \sigma_F^{(0)}(s) + \sigma_B^{(0)}(s) = 
	\frac{4\pi\alpha^2}{3s} N_{C,f} \sqrt{1-\frac{4m_f^2}{s}} \left[ \left( 1 + \frac{2m_f^2}{s} \right) G_1(s)
- 6 \frac{m_f^2}{s} G_2(s) \right], \end{eqnarray} 
with $m_f$ the final state fermion mass, $m_f \equiv m_\mu$. $N_{C,f}$ is the number of colors of the final state fermion, 
with $N_{C,f} = 1$ in the present case. $s$ 
is the cms energy, and the effective couplings $G_i(s)|_{i=1...3}$ read
\begin{eqnarray} 
G_1(s) &=& G_{1,1} + G_{1,2} + G_{1,3} = Q_e^2 Q_f^2 + 2 Q_e Q_f v_e v_f {\sf Re}[\chi_Z(s)]
          +(v_e^2+a_e^2)(v_f^2+a_f^2)|\chi_Z(s)|^2,\\ 
G_2(s) &=& (v_e^2+a_e^2) a_f^2 |\chi_Z(s)|^2, 
\\ 
G_3(s) &=& G_{3,1} + G_{3,2} = 2 Q_e Q_f a_e a_f {\sf Re}[\chi_Z(s)] + 4 v_e v_f a_e a_f |\chi_Z(s)|^2.
\end{eqnarray}
For later use we define
\begin{alignat}{8} 
\sigma_T^{\gamma\gamma} &=&  F_1~G_{1,1},~~~~&\sigma_T^{\gamma Z}     &=&  \sigma_T^{Z \gamma} = \frac{1}{2} 
F_1~G_{1,2},~~~~&\sigma_T^{Z Z}          &=&  F_1~G_{1,3} - 6 F_3~G_2
\\
\sigma_{FB}^{\gamma Z}     &=&  \sigma_{FB}^{Z \gamma} = \frac{1}{2} F_2~G_{3,1},~~~~&\sigma_{FB}^{Z Z}          &=&  
F_2~G_{3,2}. & & 
&
\end{alignat}
with
\begin{eqnarray}
F_1 &=&  \frac{4\pi\alpha^2}{3s} N_{C,f} \sqrt{1-\frac{4m_f^2}{s}} \left( 1 + \frac{2m_f^2}{s} \right),
~~~F_2 =  \frac{\pi\alpha^2}{s} N_{C,f} \left(1-\frac{4m_f^2}{s}\right),
\nonumber\\ 
F_3 &=& \frac{4\pi\alpha^2}{3s} N_{C,f} 
\left(1-\frac{4m_f^2}{s}\right) \frac{m_f^2}{s}. 
\end{eqnarray}
The reduced $Z$--propagator is given by
\begin{eqnarray} \chi_Z(s) = \frac{s}{s-M_Z^2 + i M_Z \Gamma_Z}, \end{eqnarray}
where $M_Z$ and $\Gamma_Z$ are the mass and the width of the $Z$ boson. $Q_{e,f}$ are the electromagnetic charges of the 
electron $(Q_e = -1)$ and the final state fermion, respectively, and the electro--weak couplings $v_i$ and $a_i$ read
\begin{alignat}{3} 
v_e &= \frac{1}{\sin\theta_w \cos\theta_w}\left[I^3_{w,e} - 2 Q_e \sin^2\theta_w\right],~~~~~~~ 
& a_e &= \frac{1}{\sin\theta_w \cos\theta_w} I^3_{w,e}, 
\\ 
v_f &= \frac{1}{\sin\theta_w \cos\theta_w}\left[I^3_{w,f} - 2 Q_f 
\sin^2\theta_w\right],~~~~~~~ & a_f &= \frac{1}{\sin\theta_w \cos\theta_w} I^3_{w,f}~. \end{alignat}
$\theta_w$ denotes the weak mixing angle, and $I^3_{w,i} = \pm 1/2$ the third component of the weak isospin for up and down 
particles, respectively.

When accounting for initial-state-radiation,
one obtains the following representation of the $A_{FB}$ introducing the radiators $H_{e}^{LL}$ and $H_{FB}^{LL}$ 
\cite{Beenakker:1989km}, using the notation in \cite{Bohm:1989pb},
\begin{eqnarray}
	A_{\text{FB}}(s) &=& \frac{1}{\sigma_{T}(s)} \int\limits_{z_0}^{1} {\rm d}z \, \frac{4z}{(1+z)^2} 
\tilde{H}_e^{LL}(z)
\sigma_{FB}^{(0)}(z s),~~~~~~
	\sigma_{T}(s) = \int\limits_{z_0}^{1} {\rm d}z \, H_{e}(z) \sigma_{T}^{(0)}(z s) \,, 
\nonumber\\ 
\end{eqnarray}
with
\begin{eqnarray}
\tilde{H}_e^{LL}(z) = 
\left[ H_{e}^{LL}(z) 
+ H_{FB}^{LL}(z)  \right].
\end{eqnarray}

The normalization factor $\sigma_{T}(s)$ will be calculated considering all corrections derived in 
Ref.~\cite{Ablinger:2020qvo}. We will consider different values for the threshold $z_0$. Like in 
Refs.~\cite{Blumlein:2019srk,Blumlein:2019pqb, Blumlein:2020jrf} $z_0$ is chosen as $z_0 = 4 m_\tau^2/s$, with $m_\tau$ the 
mass of the $\tau$ lepton. Another choice will be $z_0 =0.99$ or $0.999$ in accordance with \cite{Janot:2015gjr}. The variable 
$z$ is given by  $z = s'/s$, where $s'$ denotes the virtuality of the gauge boson $\gamma^*$ or $Z^*$. Furthermore, $H_{e}(z)$ 
is the radiator 
in the inclusive case, given in Ref.~\cite{Ablinger:2020qvo} and $H_{FB}^{LL}(z)$ denotes the leading-log radiator in the 
angular dependent case up to $O(\alpha^6 L^6)$,
\begin{eqnarray}
	H_{FB}^{LL}(z) = \int\limits_{0}^{1} {\rm d} x_1 \int\limits_{0}^{1} {\rm d} x_2 \left( \frac{(1+z)^2}{(x_1+x_2)^2} - 
	1 \right) \Gamma_{ee}^{LL}(x_2) \Gamma_{ee}^{LL}(x_1) \delta(x_1 x_2 - z). \label{eq:conv}
\end{eqnarray}
Here $\Gamma_{ee}^{LL}(x)$ is the leading-log operator matrix element. The product $\Gamma_{ee}^{LL}(x_1) 
\Gamma_{ee}^{LL}(x_2)$ is consistently expanded up to $O(\alpha^6 L^6)$. The latter radiator does only contain universal 
contributions. The radiators obey the expansion
\begin{eqnarray} 
H_{i}(z) &=& \delta_{ie} \delta(1-z) + \sum_{k=1}^\infty a^k \sum_{l=0}^k L^{l} H^{(k,l)}_i(z),~~~i = e, 
FB, 
\\ 
H_{i}^{LL}(z) &=& \delta_{ie} \delta(1-z) + \sum_{k=1}^\infty (a L)^k H^{(k),LL}_i(z), 
\end{eqnarray}
where $a = \alpha/(4\pi)$ and $\alpha$ denotes the fine structure constant.

We proceed in the following way in order to evaluate Eq.~(\ref{eq:conv}) analytically. First, we calculate the Mellin 
transform
\begin{eqnarray}
	\mathcal{M}[H_{FB}^{LL}(z)](n) &=& \int\limits_{0}^{1} {\rm d}z z^n H_{FB}^{LL}(z) = \int\limits_{0}^{1} {\rm d}x_1 
	\int\limits_{0}^{1} {\rm d}x_2 x_1^n x_2^n \left( \frac{(1+x_1x_2)^2}{(x_1+x_2)^2} - 1 \right) \Gamma_{ee}^{LL}(x_2) 
	\Gamma_{ee}^{LL}(x_1)
	~. \nonumber \\
\end{eqnarray}
Since this integral is not suited to be integrated with the package \texttt{HarmonicSums} directly, we compute the generating 
function
\begin{eqnarray} \label{eq:GEN}
	\mathcal{G}[H_{FB}^{LL}(z)](t) &=& \sum\limits_{n=0}^{\infty} t^n \mathcal{M}[H_{FB}^{LL}(z)](n) \nonumber\\ &=& 
\int\limits_{0}^{1} {\rm d}x_1 \int\limits_{0}^{1} {\rm d}x_2 \frac{1}{1-t x_1 x_2}
	\left( \frac{(1+x_1x_2)^2}{(x_1+x_2)^2} - 1 \right) \Gamma_{ee}^{LL}(x_2) \Gamma_{ee}^{LL}(x_1)
	~,
\end{eqnarray}
which resums the Mellin-kernel into a denominator which can be easily integrated over. After the integration over $x_1$ and 
$x_2$ we are left with generalized iterated integrals evaluated at argument $z=1$ which contain the parameter $t$ in their 
letters. We use differential equations to pull this parameter into the argument. This is straightforward since the limit $t 
\to 0$ always exists and is easily expressed in terms of known constants. Afterwards we can use the {\tt HarmonicSums} command 
\texttt{GetMoment} to get the $n$-space expression and \texttt{GeneralInvMellin} to arrive at the final result. The radiators 
are consistently expanded in $(a L)$ to $O((a L)^6)$ and are expressed using the variable $\sqrt{z}$ rather than $z$ to obtain 
a unique representation concerning the contributing iterated integrals.

At $O(\alpha)$ the complete QED initial state corrections \cite{Bardin:1989cw,Bardin:1990de,BP} are known and are given by
\begin{eqnarray} 
	A_{\text{FB}}^{(1)}(s) &=& \frac{1}{\sigma_{T}(s)} 
\int\limits_{z_0}^{1} {\rm d}z \, \frac{4z}{(1+z)^2} a(s) [\tilde{H}_e^{(1),LL}(z) L + \tilde{H}_e^{(1,0)}(z)] \sigma_{T}^{(0)}(z 
s) \,,
\end{eqnarray} 
and
\begin{eqnarray} 
{H}_e^{(1),LL}(z) &=&  4 \left[\frac{1+z^2}{1-z}\right]_+ 
\\
\tilde{H}_e^{(1,0)}(z) &=&4 \left[ - \left[\frac{1+z^2}{1-z}\right]_+  + \left(2 \zeta_2 - \frac{1}{2}\right) \delta(1-z) 
+ 
\frac{1+z^2}{1-z}
\left[2 \ln(1+z) - 2 \ln(2) -\ln(z)\right]\right]. 
\nonumber\\
\end{eqnarray} 

\section{The leading-log radiators \boldmath $H_{FB}^{LL}$} 
\label{sec:3}

\vspace*{1mm} 
\noindent 
We obtain the following leading-log radiators  $H_{FB}^{(k),LL}$ :
\begin{eqnarray} 
H_{FB}^{(1),LL}(z) &=& 0 \\
H_{FB}^{(2),LL}(z) &=&
 \frac{2 (1-z) (1+z)^2}{z}
+2 \pi \frac{(1-z)^2}{\sqrt{z}}
-8 (1+z) H_0
-8 (1-z)^2 \frac{H_{\{4,0\}}}{\sqrt{z}} \\ 
H_{FB}^{(3),LL}(z) &=& -\frac{16 (1-z) \big(4+11 z+4 
 z^2\big)}{3 z} -\pi \biggl[
         \frac{4\big(2-3 z-2 z^2-3 z^3+2 z^4\big)}{3 z^{3/2}} \nonumber \\ && +\frac{4 (1-z) (1+5 z)}{\sqrt{z}} H_0 +\frac{16 
        (1-z)^2}{\sqrt{z}} H_{\{4,1\}} \biggr] +\biggl[
         \frac{4 (1+z) \big(5-18 z-19 z^2\big)}{3 z} \nonumber \\ && -\frac{16 (1-z) (1-7 z)}{\sqrt{z}} H_{\{4,0\}}
        -96 (1+z) H_{\{4,1\}} \biggr] H_0
-8 (1+z) H_0^2 \nonumber \\ && +\biggl[
         \frac{16 (1-z) (1+z)^2}{z} -\frac{64 (1-z)^2}{\sqrt{z}} H_{\{4,0\}} \biggr] H_1 +\biggl[ \frac{16\big(2-3 z-2 z^2-3 
         z^3+2 z^4\big)}{3z^{3/2}}
\nonumber\\ &&
        +\frac{64 (1-z)^2}{\sqrt{z}} H_{\{4,1\}} \biggr] H_{\{4,0\}} +\biggl[ -\frac{16 (1-z) (1+z)^2}{z}
        +64 (1+z) H_0 \nonumber\\ && +\frac{64 (1-z)^2}{\sqrt{z}} H_{\{4,0\}} \biggr] H_{-1}
-64 (1+z) H_{0,1}
+ \frac{32 (1-z)^2}{\sqrt{z}} H_{0,\{4,0\}}
+96 (1+z) H_{0,\{4,1\}} \nonumber \\ &&
+ \frac{64 (1-z)^2}{\sqrt{z}} H_{1,\{4,0\}}
- \frac{64 (1-z)^2}{\sqrt{z}} H_{\{4,0\},\{4,1\}}
-64 (1+z) H_{-1,0} \nonumber \\ &&
- \frac{64 (1-z)^2}{\sqrt{z}} H_{-1,\{4,0\}}
+20 (1+z) \zeta_2, 
\end{eqnarray} 
with $H_{\vec{w}} \equiv H_{\vec{w}}(\sqrt{z})$ and $\zeta_k, k \in \mathbb{N}, k \geq 2$ are the values of Riemann's $\zeta$ 
function at integer argument. The iterated integrals are given by
\begin{eqnarray} 
H_{b,\vec{a}}(z) = \int_0^z dy f_b(y) H_{\vec{a}}(y),~~~H_\emptyset = 1. 
\end{eqnarray}
The indices $b, a_i$ refer to the letters in the following alphabet
\begin{eqnarray} 
\label{eq:ALPH} 
\mathfrak{A} = \left\{f_0 = \frac{1}{z}, f_1 = \frac{1}{1-z}, f_{-1} = \frac{1}{1+z}, 
f_{\{4,0\}} = \frac{1}{1+z^2}, f_{\{4,1\}} = \frac{z}{1+z^2} \right\}, 
\end{eqnarray}
where the last two letters in (\ref{eq:ALPH}) are cyclotomic. The function $H_{FB}^{(2),LL}(z)$ has been calculated in 
\cite{Beenakker:1989km} in an equivalent representation. The radiators $H_{FB}^{(4-6),LL}(z)$ are too voluminous to be 
presented here and are given in computer-readable form in the attachment.

We finally illustrate the structure of the radiators expanding them for their leading small and a few large $z$ terms. One 
obtains
\begin{eqnarray} 
zH_{FB}^{(2),LL}(z) &\simeq& 2 + O(\sqrt{z}) \\ 
zH_{FB}^{(3),LL}(z) &\simeq& - \frac{8 \pi}{3 \sqrt{z}} + 
\frac{10}{3} \ln(z) - \frac{32}{3}
+ O(\sqrt{z} \ln(z)) 
\\ 
zH_{FB}^{(4),LL}(z) &\simeq& 
\frac{32}{9 z}
- \frac{11}{6} \ln^2(z) 
+ \frac{10}{27} \ln(z)
-38
-43 \zeta_2
\nonumber \\ &&
+ \frac{\pi}{\sqrt{z}} 
\biggl( 
	  \frac{4}{3} \ln(z) 
	+ \frac{328}{27} 
\biggr)
+ O(\sqrt{z} \ln^2(z)) 
\\ 
zH_{FB}^{(5),LL}(z) &\simeq& 
-\frac{512}{81 z} 
+\frac{5}{36} \ln^3(z) 
+\frac{97}{18} \ln^2(z)
+\biggl[
    -\frac{299}{27} 
	+\frac{215 \zeta_2}{18} 
\biggr] \ln(z)
+ \frac{72283}{243}
\nonumber \\ &&
+ \frac{2551}{27} \zeta_2
+ \frac{320}{9} \zeta_3
+ \frac{\pi}{\sqrt{z}}
\biggl(
	\frac{2}{3} \ln^2(z)
	+ \frac{92}{27} \ln(z)
	+ \frac{460}{81}
\nonumber \\ &&
	+ \frac{4}{9} \zeta_2
\biggr)
+ O(\sqrt{z} \ln^3(z)) 
\\ 
zH_{FB}^{(6),LL}(z) &\simeq& 
- \frac{10240}{729 z}
- \frac{256}{81 z} \ln(z)
+ \frac{97}{720} \ln^4(z)
+ \frac{79}{162} \ln^3(z)
+ \biggl(
	\frac{17059}{2430}
	- \frac{19}{180} \zeta_2
\biggr) \ln^2(z)
\nonumber \\ &&
+ \biggl(
	- \frac{159109}{7290}
	- \frac{2825}{81} \zeta_2
	+ \frac{136}{9} \zeta_3
\biggr) \ln(z)
- \frac{2459744}{10935}
+ \frac{87173}{1215} \zeta_2
- \frac{53216}{405} \zeta_3
\nonumber \\ &&
+ \frac{5659}{300} \zeta_2^2
+ \frac{\pi}{\sqrt{z}}
\biggl(
	- \frac{1}{18} \ln^3(z)
	- \frac{32}{9} \ln^2(z)
	+ \biggl[
		- \frac{39212}{1215}
		- \frac{73}{9} \zeta_2
	\biggr] \ln(z)
\nonumber \\ &&
	- \frac{401806}{3645}
	- \frac{15904}{405} \zeta_2
	- \frac{128}{9} \zeta_3
\biggr)
+ O(\sqrt{z} \ln^4(z)). 
\end{eqnarray} 
For the large $z$ representation we set $u = 1 - \sqrt{z}$ and obtain
\begin{eqnarray} 
H_{FB}^{(2),LL}(z) &\simeq& 32 u - 16 u^2 + \frac{112}{3} u^3 + \frac{56}{3} u^4 + O(u^5) 
\\
H_{FB}^{(3),LL}(z) &\simeq& \frac{64}{3} u -\frac{416}{3} u^2 +\frac{1408}{9} u^3 -\frac{736}{9} u^4 +\Biggl(
	-256 u
	+128 u^2
	-\frac{896}{3} u^3
	-\frac{448}{3} u^4
\Biggr) \nonumber\\ && \times \ln(2u)  
+ O(u^5) 
\\ 
H_{FB}^{(4),LL}(z) &\simeq& 
  \frac{27328}{27} u
- \frac{13664}{27} u^2
+ \frac{162080}{81} u^3
+ \frac{36112}{81} u^4
+ \biggl(
	 1024 u
	- 512 u^2
\nonumber\\ &&
	+ \frac{3584}{3} u^3
	+ \frac{1792}{3} u^4
\biggr) \left( \ln^2(2u) - \zeta_2 \right)
+ \biggl(
	1024 
	- \frac{3328}{3} u
	+ \frac{2176}{3} u^2
\biggr) 
\nonumber\\ && \times u^2 \ln(2u)
+ O(u^5) 
\\ 
H_{FB}^{(5),LL}(z) &\simeq& 
\frac{387712}{81} u
- \frac{523328}{81} u^2
+ \frac{2954144}{243} u^3
- \frac{328688}{243} u^4
+ \biggl(
	\frac{2048}{3}
	+ \frac{11264}{3} u
\nonumber\\ &&
	- \frac{34304}{9} u^2
	+ \frac{28928}{9} u^3
\biggr) u \left( \zeta_2 - \ln^2(2u) \right)
+ \biggl(
	- \frac{8192}{3}
	+ \frac{4096}{3} u
	- \frac{28672}{9} u^2
\nonumber\\ &&
	- \frac{14336}{9} u^3
\biggr) u \left( 2\zeta_3 -3 \zeta_2 \ln(2u) + \ln^3(2u) \right)
+ \biggl(
	- \frac{73216}{9}
	+ \frac{30464}{9} u
\nonumber\\ &&
	- \frac{419840}{27} u^2
	- \frac{108544}{27} u^3
\biggr) u \ln(2u) 
+ O(u^5) 
\\ 
H_{FB}^{(6),LL}(z) &\simeq& 
\frac{153977536}{3645} u
- \frac{136846688}{3645} u^2
+ \frac{1140995264}{10935} u^3
+ \frac{60397216}{10935} u^4
+ \biggl(
  - \frac{894976}{27}
\nonumber\\ &&
  + \frac{300032}{27} u
  - \frac{24477184}{405} u^2
  - \frac{7326464}{405} u^3
\biggr) u \left( \zeta_2 - \ln^2(2u) \right)
+ \biggl(
	\frac{16384}{5}
  - \frac{8192}{5} u
\nonumber\\ &&
  + \frac{57344}{15} u^2
  + \frac{28672}{15} u^3
\biggr) u \left( \zeta_2^2 + \frac{5}{3} \ln^4(2u) \right)
+ \biggl(
	\frac{32768}{9}
  + \frac{81920}{9} u
  - \frac{1132544}{135} u^2
\nonumber\\ &&
  + \frac{1276928}{135} u^3
\biggr) u \left( 2\zeta_3 -3 \zeta_2 \ln(2u) + \ln^3(2u) \right)
+ \biggl(
  -\frac{886784}{27}
  +\frac{1338368}{27} u
\nonumber\\ &&
  -\frac{35939456}{405} u^2
  +\frac{5414336}{405} u^3
  - \biggl[
	  -\frac{131072}{3}
	  +\frac{65536}{3} u
	  -\frac{458752}{9} u^2
\nonumber\\ &&
	  -\frac{229376}{9} u^3
  \biggr] \zeta_3
\biggr) u \ln(2u)
+ \biggl(
  - 32768
  + 16384 u
  -\frac{114688}{3} u^2
  -\frac{57344}{3} u^3
\biggr) 
\nonumber\\ &&
\times u \zeta_2 \ln^2(2u)
+ O(u^5). 
\end{eqnarray} 
These terms are the beginning of the respective series used in the numerical representation to which we turn now.

\section{Numerical results} \label{sec:4}

\vspace*{1mm} 
\noindent
For the numerical evaluation of the new radiators we calculated 60 terms 
of the expansions around $\sqrt{z}=0$ and $\sqrt{z}=1$
using {\tt HarmonicSums}. The constants associated with cyclotomic harmonic polylogarithms for cyclotomy {\sf c=4}, which 
are needed 
for the 
expansion around $\sqrt{z}=1$ are known at least up to weight {\sf w = 5} \cite{Ablinger:2011te}.
However, all constants except of $\pi$ cancel in the final results.
We switch between these two expansions at $z=1/4$, where the absolute difference of the respective approximation is always 
smaller than $10^{-12}$. The usual harmonic polylogarithms are calculated using the implementations in \cite{Gehrmann:2001pz, 
Ablinger:2018sat}. In the numerical illustration we refer to the electroweak parameters given in \cite{PDG}.

\begin{figure}[H]
	\centering
	\includegraphics[width=0.8\textwidth]{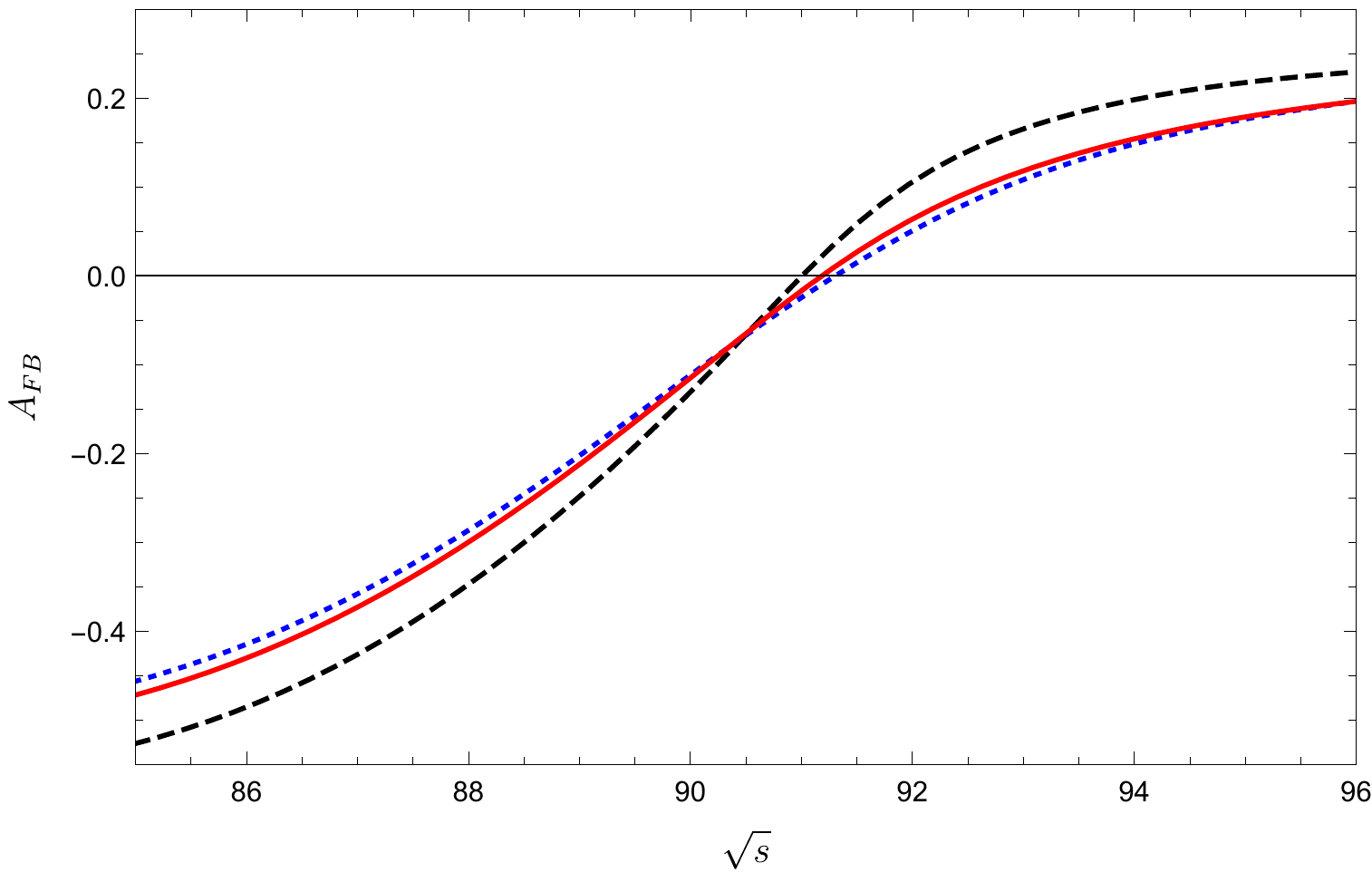}
	\caption{\sf $A_{FB}$ and its initial state QED corrections as a function of $\sqrt{s}$. Black (dashed)
	the Born approximation, blue (dotted) the $O(\alpha)$ improved
	approximation, red (full) also including the leading-log 
	improvement up to $O(\alpha^6)$  for $s'/s \geq 4 m_\tau^2/s$.}
\label{fig:1}
\end{figure}

In Figure~\ref{fig:1} we illustrate the effect of the initial--state QED corrections on the forward--backward asymmetry for 
a wider range of cms energies around the $Z$-peak with $\sqrt{s} \in [85,96]~\GeV$ to provide an overall impression
on the size of the radiative corrections, which will be detailed below.

Let us define
\begin{eqnarray}
	\Delta A_{FB} = 1 - \frac{A_{FB}^{(l)}}{A_{FB}^{(0)}},
	\label{eq:plot2}
\end{eqnarray}
\begin{figure}[H]
	\centering
	\includegraphics[width=0.8\textwidth]{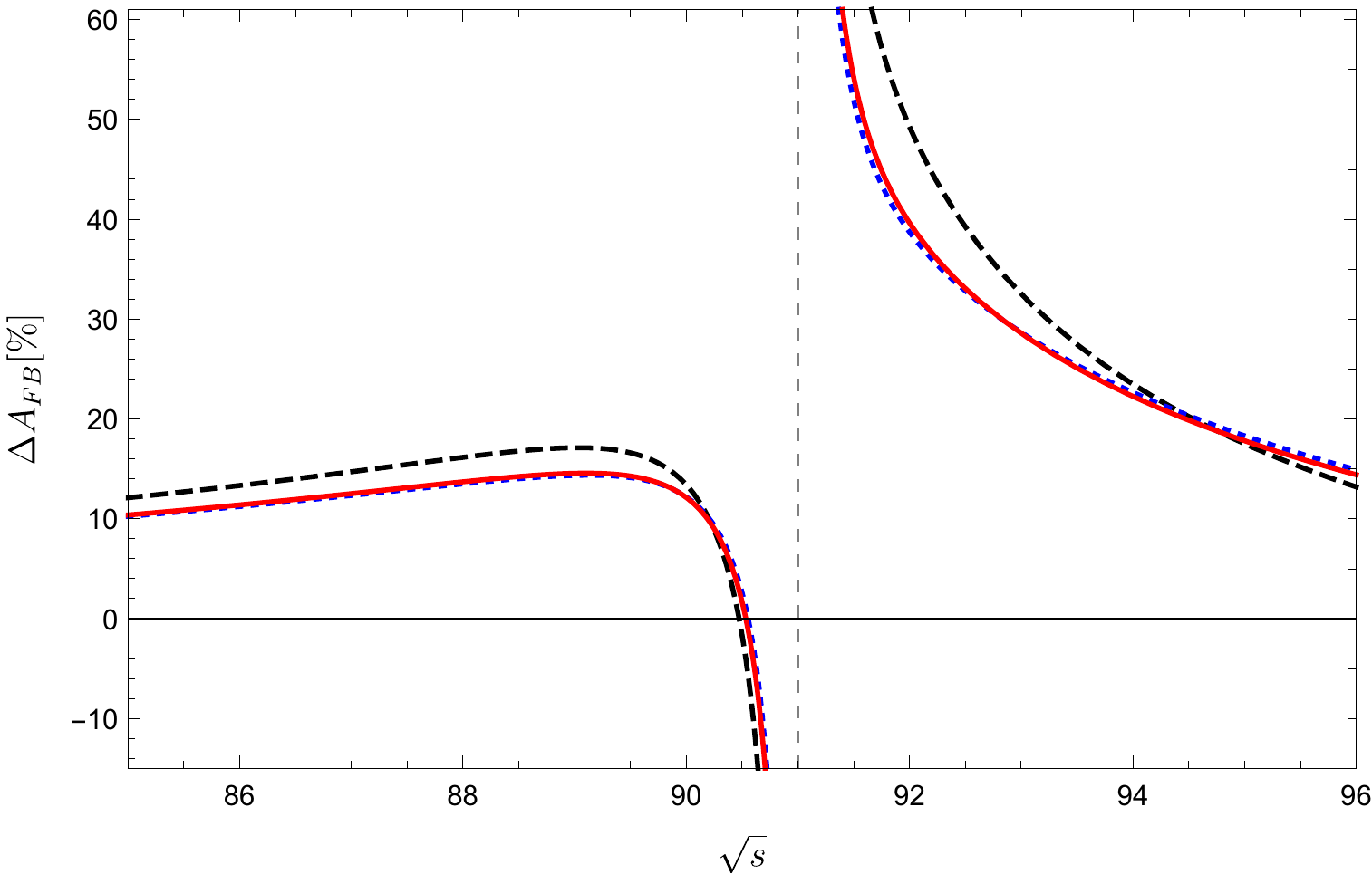}
	\caption{\sf $\Delta A_{FB}$ in \% as defined in Eq.~(\ref{eq:plot2}) 
	as a function of $\sqrt{s}$. 
	Black (dashed) the $O(\alpha)$ improved approximation, 
	blue (dotted) the $O(\alpha^2 L^2)$ improved approximation, 
	red (full) also including the leading-log 
	improvement up to $O(\alpha^6)$ for $s'/s \geq 4 m_\tau^2/s$.}
\label{fig:2}
\end{figure}
\noindent
where $l$ denotes the order to which the initial state radiation
is taken into account. $l=0$ corresponds to the Born approximation. In Figure~\ref{fig:2} we illustrate the relative size 
of the ISR corrections to the Born cross section.

In Table~\ref{tab:AFB_1} we illustrate the effect of the different orders of the QED initial state corrections of the 
forward--backward asymmetry at the $Z$-peak and two more values of $s = s_\pm$ 
\cite{Janot:2015gjr}. The ISR corrections are not monotonic order by order, even leading to a sign change in the 
corrected  values of $A_{FB}$ at $s = M_Z^2$. Here the ISR corrections are largest. 

\noindent 
If we compare the known one-loop 
corrected values of $A_{FB}$ with the highest radiative correction calculated in this paper we obtain further corrections
of $-3$ \% for $s_-$ and  $-1 \%$ for $s_+$. These values supersede previous estimates, based on assumptions, in 
\cite{Janot:2015gjr}. 
\begin{table}[H]
	\centering
	\begin{tabular}{rrrr}
		\toprule
		 				& $A_{FB}(s_-)$ 	& $A_{FB}(M_Z^2)$ & $A_{FB}(s_+)$
		\\ \midrule
		$\mathcal{O}(\alpha^0)$		& $-0.3564803$		& $ 0.0225199$	& $0.2052045$
		\\
		$\mathcal{O}(\alpha L^1)$	& $-0.2945381$		& $-0.0094232$	& $0.1579347$
		\\
		$\mathcal{O}(\alpha L^0)$	& $-0.2994478$		& $-0.0079610$	& $0.1611962$
		\\
		$\mathcal{O}(\alpha^2 L^2)$	& $-0.3088363$		& $ 0.0014514$	& $0.1616887$
		\\
		$\mathcal{O}(\alpha^3 L^3)$	& $-0.3080578$		& $ 0.0000198$	& $0.1627252$
		\\
		$\mathcal{O}(\alpha^4 L^4)$	& $-0.3080976$		& $ 0.0001587$	& $0.1625835$
		\\	
		$\mathcal{O}(\alpha^5 L^5)$	& $-0.3080960$		& $ 0.0001495$	& $0.1625911$
		\\ 
		$\mathcal{O}(\alpha^6 L^6)$	& $-0.3080960$		& $ 0.0001499$	& $0.1625911$
		\\ \bottomrule
	\end{tabular}
	\caption{$A_{FB}$ evaluated at $s_-=(87.9\,{\rm GeV})^2$, $M_Z^2$
			and $s_+=(94.3\,{\rm GeV})^2$ for the cut $z>4m_\tau^2/s$.}
	\label{tab:AFB_1}
\end{table}

If cuts of $s'$ are applied the ISR corrections turn out to be smaller, as illustrated in Tables~\ref{tab2} 
and \ref{tab3}.
As noted in \cite{Bohm:1989pb} also an approximative treatment of the forward--backward asymmetry
\begin{eqnarray}
	A_{\text{FB}}(s) & \approx & \frac{1}{\sigma_{T}(s)} \int\limits_{z_0}^{1} {\rm d}z \, \frac{4z}{(1+z)^2} H_{e}(z) 
\sigma_{FB}^{(0)}(z s) \,, 
\end{eqnarray}
has been used in the literature \cite{Riemann:1989ca} close to the $Z$-peak, where $H_e(z)$ denotes the inclusive 
\begin{table}[H]
	\centering
	\begin{tabular}{rrrr}
		\toprule
		 				& $A_{FB}(s_-)$ 	& $A_{FB}(M_Z^2)$ & $A_{FB}(s_+)$
		\\ \midrule
		$\mathcal{O}(\alpha^0)$		& $-0.2714847$		& $0.01545540$	& $0.2539839$
		\\
		$\mathcal{O}(\alpha L^1)$	& $-0.2913715$		& $0.01244012$	& $0.2787800$
		\\
		$\mathcal{O}(\alpha L^0)$	& $-0.2923065$		& $0.01266498$	& $0.2794039$
		\\
		$\mathcal{O}(\alpha^2 L^2)$	& $-0.2922610$		& $0.01489549$	& $0.2760952$
		\\
		$\mathcal{O}(\alpha^3 L^3)$	& $-0.2917702$		& $0.01430419$	& $0.2764488$
		\\
		$\mathcal{O}(\alpha^4 L^4)$	& $-0.2918380$		& $0.01439578$	& $0.2763865$
		\\	
		$\mathcal{O}(\alpha^5 L^5)$	& $-0.2918313$		& $0.01438608$	& $0.2763934$
		\\ 
		$\mathcal{O}(\alpha^6 L^6)$	& $-0.2918318$		& $0.01438681$	& $0.2763929$
		\\ \bottomrule
	\end{tabular}
	\caption{$A_{FB}$ evaluated at $s_-=(87.9\,{\rm GeV})^2$, $M_Z^2$
			and $s_+=(94.3\,{\rm GeV})^2$ for the cut $z>0.99$.}
	\label{tab2}
\end{table}

\noindent
radiator 
appearing in case of $\sigma_T(s)$ \cite{Ablinger:2020qvo}. In the sub--leading corrections to the radiators 
\cite{Ablinger:2020qvo}, Wilson coefficients appear, which are different for $\sigma_T$ and $\sigma_{FB}$. Therefore the 
radiator for $\sigma_T$ cannot be used in its sub--leading terms for this reason. The numerical results in 
Table~\ref{tab4}--\ref{tab6} show, that contributions due to the angular radiators are indeed suppressed if measuring 
at the $Z$-peak w.r.t. to the inclusive ones also in higher orders. This holds quite irrespectively of the cuts in $s'$.
\begin{table}[H]
	\centering
	\begin{tabular}{rrrr}
		\toprule
		 				& $A_{FB}(s_-)$ 	& $A_{FB}(M_Z^2)$ & $A_{FB}(s_+)$
		\\ \midrule
		$\mathcal{O}(\alpha^0)$		& $-0.2683019$		& $0.01538044$	& $0.2576633$
		\\
		$\mathcal{O}(\alpha L^1)$	& $-0.2905494$		& $0.01623701$	& $0.2797643$
		\\
		$\mathcal{O}(\alpha L^0)$	& $-0.2913658$		& $0.01630129$	& $0.2805232$
		\\
		$\mathcal{O}(\alpha^2 L^2)$	& $-0.2899912$		& $0.01655798$	& $0.2787005$
		\\
		$\mathcal{O}(\alpha^3 L^3)$	& $-0.2898133$		& $0.01641903$	& $0.2787137$
		\\
		$\mathcal{O}(\alpha^4 L^4)$	& $-0.2898401$		& $0.01645212$	& $0.2786941$
		\\	
		$\mathcal{O}(\alpha^5 L^5)$	& $-0.2898362$		& $0.01644636$	& $0.2786984$
		\\ 
		$\mathcal{O}(\alpha^6 L^6)$	& $-0.2898368$		& $0.01644713$	& $0.2786978$
		\\ \bottomrule
	\end{tabular}
	\caption{$A_{FB}$ evaluated at $s_-=(87.9\,{\rm GeV})^2$, $M_Z^2$
			and $s_+=(94.3\,{\rm GeV})^2$ for the cut $z>0.999$.}
	\label{tab3}
\end{table}
\begin{table}[H]
	\centering
	\begin{tabular}{rrrrrrr}
		\toprule
		 		& all $(s_-)$ 	& ang. $(s_-)$ 
				& all $(M_Z^2)$ & ang. $(M_Z^2)$ 
				& all $(s_+)$ 	& ang. $(s_+)$ 
		\\ \midrule
		$\mathcal{O}(\alpha^2 L^2)$	& $-9 \cdot 10^{-3}$ & $-9 \cdot 10^{-5}$ & $9 \cdot 10^{-3}$ & $-1  \cdot 10^{-5}$ & $5 \cdot 10^{-4}$ & $-6 \cdot 10^{-5}$
		\\
		$\mathcal{O}(\alpha^3 L^3)$	& $8 \cdot 10^{-4}$ & $-4 \cdot 10^{-6}$ & $-1 \cdot 10^{-3}$ & $-8 \cdot 10^{-7}$ & $1 \cdot 10^{-3}$ & $-3 \cdot 10^{-6}$
		\\
		$\mathcal{O}(\alpha^4 L^4)$	& $-4 \cdot 10^{-5}$ & $-3 \cdot 10^{-7}$ & $1 \cdot 10^{-4}$ & $-7 \cdot 10^{-8}$ & $-1 \cdot 10^{-4}$ & $-7 \cdot 10^{-8}$
		\\	
		$\mathcal{O}(\alpha^5 L^5)$	& $2 \cdot 10^{-6}$ & $-1 \cdot 10^{-8}$ & $-9 \cdot 10^{-6}$ & $-4  \cdot 10^{-9}$ & $8 \cdot 10^{-6}$ & $1 \cdot 10^{-8}$
		\\ 
		$\mathcal{O}(\alpha^6 L^6)$	& $-6 \cdot 10^{-8}$ & $-6 \cdot 10^{-10}$ & $4 \cdot 10^{-7}$ & $-1 \cdot 10^{-10}$ & $-3 \cdot 10^{-8}$ & $2 \cdot 10^{-9}$
		\\ \bottomrule
	\end{tabular}
	\caption{Contribution to $A_{FB}$ at different orders of $\alpha L$ evaluated 
			at $s_-=(87.9\,{\rm GeV})^2$, $M_Z^2$
			and $s_+=(94.3\,{\rm GeV})^2$ for the cut $z>4m_\tau^2/s$.
			all: full contributions,  ang.: only the angular dependent contributions.}
	\label{tab4}
\end{table}
\begin{table}[H]
	\centering
	\begin{tabular}{rrrrrrr}
		\toprule
		 		& all $(s_-)$ 	& ang. $(s_-)$ 
				& all $(M_Z^2)$ 	& ang. $(M_Z^2)$ 
				& all $(s_+)$ 	& ang. $(s_+)$ 
		\\
\midrule
		$\mathcal{O}(\alpha^2 L^2)$	& $5 \cdot 10^{-5}$ & $-5 \cdot 10^{-8}$ & $2 \cdot 10^{-3}$ & $-2  \cdot 10^{-9}$ & $-3 \cdot 10^{-3}$ & $5 \cdot 10^{-8}$
		\\
		$\mathcal{O}(\alpha^3 L^3)$	& $5 \cdot 10^{-4}$ & $-3 \cdot 10^{-8}$ & $-6 \cdot 10^{-4}$ & $-1 \cdot 10^{-9}$ & $4 \cdot 10^{-4}$ & $3 \cdot 10^{-8}$
		\\
		$\mathcal{O}(\alpha^4 L^4)$	& $-7 \cdot 10^{-5}$ & $-9 \cdot 10^{-9}$ & $9 \cdot 10^{-5}$ & $-3 \cdot 10^{-10}$ & $-6 \cdot 10^{-5}$ & $9 \cdot 10^{-9}$
		\\	
		$\mathcal{O}(\alpha^5 L^5)$	& $7 \cdot 10^{-6}$ & $-2 \cdot 10^{-9}$ & $-1 \cdot 10^{-5}$ &$-4  \cdot 10^{-11}$ & $7 \cdot 10^{-6}$ & $2 \cdot 10^{-9}$
		\\ 
		$\mathcal{O}(\alpha^6 L^6)$	& $-5 \cdot 10^{-7}$ & $-2 \cdot 10^{-10}$ & $7 \cdot 10^{-7}$ & $-4 \cdot 10^{-12}$ & $-5 \cdot 10^{-7}$ & $2 \cdot 10^{-10}$
		\\ \bottomrule
	\end{tabular}
	\caption{Contribution to $A_{FB}$ at different orders of $\alpha L$ evaluated 
			at $s_-=(87.9\,{\rm GeV})^2$, $M_Z^2$
			and $s_+=(94.3\,{\rm GeV})^2$ for the cut $z>0.99$.
			all: full contributions,  ang.: only the angular dependent contributions.}
	\label{tab5}
\end{table}
For completeness we also quantify the final state and initial--final interference contributions at $O(\alpha)$, cf. 
e.g.~\cite{Bohm:1989pb,BP,Bardin:1989cw,Bardin:1990de} for the respective numerator, $\sigma_{FB}$, and denominator, $\sigma_T$ 
contributions to $A_{FB} = \sigma_{FB}/\sigma_T$. The inclusive final state correction to $\sigma_{FB}$ and $\sigma_T$
imply the factors
\begin{eqnarray} 
1 + \delta_{FB}^F,~~\text{with}~~\delta_{FB}^F = 0~~~~\text{and}~~~~ 
1 + \delta_T^F,~~\text{with}~~\delta_T^F = 3 a 
\end{eqnarray}
for the $\mu^+ \mu^-$ final state. The multiplicative numerical correction to $A_{FB}$ is
$0.998149$.

The initial--final interference term \cite{BP,Bardin:1990de} has a more involved representation
compared to the final state corrections, since there are different correction terms to the different 
pieces of the angular averaged Born cross section. In particular these terms are important because
of logarithmically large terms of $\ln(1-s'_{\rm min}/s)$ in the region $s'_{\rm min}$ close to $s$
as chosen in \cite{Janot:2015gjr}. The correction term $1 + \delta_{FB}^{FI}$ modifying $A_{FB}$ to 
$O(\alpha)$ is given by 
\begin{eqnarray} 
\delta_{FB}^{FI}  &=& - \frac{1}{\sigma_T^{\rm Born}}
\Biggl\{\sigma_{FB}^{\gamma Z} 
       {\sf Re}[R_T^{\gamma \gamma} + R_T^{Z Z}] 
     + \sigma_{FB}^{Z Z} {\sf Re}[R_T^{Z Z}] \Biggr\}
\nonumber\\ &&
+ \frac{1}{A_{FB}} \Biggl\{{\sf Re}[R_{FB}^{\gamma \gamma}] \frac{\sigma_T^{\gamma \gamma}}{\sigma_T^{\rm Born}}
+ {\sf Re}[
R_{FB}^{\gamma \gamma}+R_{FB}^{Z Z}] 
\frac{\sigma_T^{\gamma Z}}{\sigma_T^{\rm Born}}
+ {\sf Re}[
R_{FB}^{Z Z}] 
\frac{\sigma_T^{Z Z}}{\sigma_T^{\rm Born}}\Biggr\},
\end{eqnarray}
with $R_{A}^{ij},~~A = T,FB,~~i,j = \gamma, Z$, the corresponding radiators \cite{BP,Bardin:1990de,Bardin:1990fu}
and
\begin{eqnarray} 
\label{eq:cut1}
\delta_{FB}^{FI} &=&   -0.0821 ~~~~~~~~\text{for}~~ \sqrt{s} = 87.9~\GeV ~~~~s'/s > 0.99
\\
\delta_{FB}^{FI} &=&   +0.0928~~~~~~~~\text{for}~~ \sqrt{s} = 94.3~\GeV ~~~~s'/s > 0.99
\\
\delta_{FB}^{FI} &=&   -0.1954~~~~~~~~\text{for}~~ \sqrt{s} = 87.9~\GeV ~~~~s'/s > 0.999
\\
\label{eq:cut2}
\delta_{FB}^{FI} &=&   +0.2223~~~~~~~~\text{for}~~ \sqrt{s} = 94.3~\GeV ~~~~s'/s > 0.999
\end{eqnarray}
The tight cuts in (\ref{eq:cut1}--\ref{eq:cut2}) imply a correction term of up to $\pm9\%$ for 
$z > 0.99$ and up to $\pm22\%$ for $z > 0.999$.

\begin{table}[H]
	\centering
	\begin{tabular}{rrrrrrr}
		\toprule
		 		& all $(s_-)$ 	& ang. $(s_-)$ 
				& all $(M_Z^2)$ 	& ang. $(M_Z^2)$ 
				& all $(s_+)$ 	& ang. $(s_+)$ 
		\\
\midrule
		$\mathcal{O}(\alpha^2 L^2)$	& $1 \cdot 10^{-3}$ & $-5 \cdot 10^{-10}$ & $3 \cdot 10^{-4}$ & $2  \cdot 10^{-11}$ & $-2 \cdot 10^{-3}$ & $5 \cdot 10^{-10}$
		\\
		$\mathcal{O}(\alpha^3 L^3)$	& $2 \cdot 10^{-4}$ & $-4 \cdot 10^{-10}$ & $-1 \cdot 10^{-4}$ & $2 \cdot 10^{-11}$ & $1 \cdot 10^{-5}$ & $4 \cdot 10^{-10}$
		\\
		$\mathcal{O}(\alpha^4 L^4)$	& $-3 \cdot 10^{-5}$ & $-2 \cdot 10^{-10}$ & $3 \cdot 10^{-5}$ & $9 \cdot 10^{-12}$ & $-2 \cdot 10^{-5}$ & $2 \cdot 10^{-10}$
		\\	
		$\mathcal{O}(\alpha^5 L^5)$	& $4 \cdot 10^{-6}$ & $-5 \cdot 10^{-11}$ & $-6 \cdot 10^{-6}$ & $3  \cdot 10^{-12}$ & $4 \cdot 10^{-6}$ & $5 \cdot 10^{-11}$
		\\ 
		$\mathcal{O}(\alpha^6 L^6)$	& $-5 \cdot 10^{-7}$ & $-1 \cdot 10^{-11}$ & $8 \cdot 10^{-7}$ & $5 \cdot 10^{-13}$ & $-6 \cdot 10^{-7}$ & $1 \cdot 10^{-11}$
		\\ \bottomrule
	\end{tabular}
	\caption{Contribution to $A_{FB}$ at different orders of $\alpha L$ evaluated 
			at $s_-=(87.9\,{\rm GeV})^2$, $M_Z^2$
			and $s_+=(94.3\,{\rm GeV})^2$ for the cut $z>0.999$.
			all: full contributions,  ang.: only the angular dependent contributions.}
	\label{tab6}
\end{table}

\appendix
\section{\boldmath $N$ space results} 
\label{sec:A}

\vspace*{1mm} 
\noindent 
In the following we are presenting the first radiators in Mellin $N$-space.
\begin{eqnarray} 
H_{FB}^{(1),LL}(N) &=& 0 
\\ 
H_{FB}^{(2),LL}(N) &=&
 \frac{8 \big(3 N^2+3 N-1\big) P_1}{(N-1) N^2 (N+1)^2 (N+2) (2N-1) (2N+3)} 
\nonumber \\ 
&& -\frac{32 \big(4 N^2+4 N-1\big) 
(-1)^N }{(2N-1) (2N+1) (2N+3)} [S_{-1} + \ln(2)] 
\\ 
H_{FB}^{(3),LL}(N) &=&
 \frac{64 \big(3 N^2+3 N-1\big) P_2}{(N-1) N^2 (N+1)^2 (N+2) (2N-1) (2N+3)} S_1 \nonumber \\ && -\frac{4 P_7}{3 (N-1)^2 N^3 
(N+1)^3 (N+2)^2 (2N-3) (2N-1)^2 (2N+3)^2 (2N+5)} \nonumber \\ && +(-1)^N \Biggl\{\frac{8 P_4}{N (N+1) (2N-1) (2N+1) (2N+3)} 
S_{-2} \nonumber \\ && -\frac{256 \big(4 N^2+4 N-1\big)}{(2N-1) (2N+1) (2N+3)} \left[S_{-1,1} - \frac{1}{2} \ln^2(2)
+ \sum_{i=1}^N \frac{\ln(2)+S_{-1}\big(i\big)}{1+2i} \right] \nonumber\\ && -\frac{128 P_5}{3 (2N-3) (2N-1)^2 (2N+1)^2 
  (2N+3)^2 (5+2 N)} S_{-1} \nonumber \\ &&
        +\frac{4 P_3}{N (N+1) (2N-1) (2N+1) (2N+3)} \zeta_2 \nonumber \\ && -\frac{128 P_6}{3 (2N-3) (2N-1)^2 (2N+1)^2 
        (2N+3)^2 (2N+5)} \ln(2)
\Biggr\}, 
\end{eqnarray}
with the polynomials
\begin{eqnarray}
P_1 &=& 4 N^4+8 N^3-N^2-5 N-3 
\\ 
P_2 &=& 4 N^4+8 N^3-N^2-5 N-3, 
\\ P_3 &=& 32 N^4+16 N^3-48 N^2-14 N+9, 
\\ 
P_4 &=& 64 N^4+80 N^3-24 N^2-22 N+9,
\\
P_5 &=& 128 N^7+256 N^6-320 N^5-144 N^4+1128 N^3
	+104 N^2-936 N+27, \\ P_6 &=& 896 N^7+2944 N^6-512 N^5-7344 N^4-2760 N^3
	+2816 N^2+36 N-243, \\ P_7 &=& 5120 N^{14}+35840 N^{13}+73856 N^{12}-22784 N^{11}-252848 N^{10}
	-201200 N^9 \nonumber\\ &&
+193136 N^8
+264800 N^7-43255 N^6-113445 N^5
	+7512 N^4+13587 N^3 \nonumber\\ &&
+1539 N^2
+8262 N-1620. 
\end{eqnarray} 
Since the expressions for the higher order radiators are voluminous, they are given in a computer-readable file attached to 
this paper. Here the functions $S_{\vec{a}} \equiv S_{\vec{a}}(N)$ denote the harmonic sums 
\cite{Vermaseren:1998uu,Blumlein:1998if}
\begin{eqnarray} 
S_{b,\vec{a}}(N) = \sum_{k=1}^N \frac{({\rm sign}(b))^k}{k^{|b|}} S_{\vec{a}}(k),~~S_\emptyset = 1,~~a,b_i 
\in \mathbb{Z} \backslash \{0\}. 
\end{eqnarray}
Despite the fact that already at 2nd order in $z$--space cyclotomic harmonic polylogarithms occur, cyclotomic harmonic sums 
are only contributing from 3rd order onward. As usual, the $N$--space expressions turn out to be structurally simpler than the 
$z$--space expressions. However, there are evanescent poles at half integer arguments $N = 1/2, 3/2, ...$. As has been seen in 
the small $z$ expansion of the radiators, they are all tractable for $N > 1$. The pure singlet kernels are introducing a 
rightmost pole at $N = 1$ and poles left to this are allowed.

One may numerically represent the higher order corrections also starting from Mellin $N$ space, as described in 
Refs.~\cite{ANCONT,Blumlein:2009ta}. For this asymptotic representations and the recurrences of the respective expressions 
need to be known, which can be easily obtained using the package {\tt HarmonicSums}. The recurrences are implied by the 
quantities itself in terms of the hierarchic sum--structures.

\vspace*{5mm} 
\noindent 
{\bf Acknowledgments.}\\ 
We would like to thank W.~Hollik, P.~Janot, P.~Marquard, G.~Passarino, and C.~Schneider for discussions. This project has received 
funding from the European Union's Horizon 2020 research and innovation programme 
under the Marie Sk\/{l}odowska-Curie grant agreement No. 764850, SAGEX, COST action CA16201: Unraveling new physics at the LHC 
through the precision frontier.
 
\end{document}